\documentclass[aps,prl,superscriptaddress,showpacs,twocolumn]{revtex4}
\usepackage{graphicx}
\usepackage{dcolumn}
\usepackage{bm}
\usepackage{graphicx}
\usepackage{amsmath, amsfonts, amssymb,mathrsfs}
\usepackage{pstricks}
\usepackage{amsxtra}
\usepackage{amsthm}
\usepackage{natbib}
\def\ie{i.e.}

\def\be{\begin{equation}}      
\def\ee{\end{equation}}
\def\beu{\begin{equation*}}   
\def\eeu{\end{equation*}}

\providecommand{\abs}[1]{\left\lvert#1\right\rvert}   
\providecommand{\ket}[1]{\left|#1\right\rangle}

\providecommand{\bra}[1]{\left\langle#1\right|}
\providecommand{\mean}[1]{\left\langle#1\right\rangle}

\renewcommand{\vec}[1]{\bm{#1}}
\providecommand{\Bext}{B_\text{ext}}
\providecommand{\Tp}{T_+}

\providecommand{\TO}{T_0}

\providecommand{\DelO}{\Delta_0}
\providecommand{\Delm}{\Delta_{-}}

\providecommand{\GamO}{\Gamma_0}
\providecommand\Dperp{D_\perp}

\providecommand\Dz{D_z}
\providecommand\Dvec{\vec{D}}

\providecommand\Sz{S_z}

\providecommand\Svec{\vec{S}}
\providecommand{\zhat}{\hat{z}}
\providecommand{\txt}{\textrm}
\newcommand{\rs}{R}
\begin{document}

\title{Dynamic Nuclear Polarization in Double Quantum Dots}

\author{M.~Gullans}
\author{J.~J.~Krich}
\affiliation{Department of Physics, Harvard University, Cambridge, MA 
02138, USA}
\author{J.~M.~Taylor}
\affiliation{Department of Physics, Massachusetts Institute of 
Technology, Cambridge, MA 02139, USA}
\affiliation{Joint Quantum Institute and the National 
Institute of Standards and Technology, College Park, Maryland 20472, USA}
\author{H.~Bluhm}
 \author{B.~I.~Halperin}
 \author{C.~M.~Marcus}
  \affiliation{Department of Physics, Harvard University, Cambridge, MA 
02138, USA}
 \author{M.~Stopa}
\affiliation{Center for Nanoscale Systems, Harvard University, 
Cambridge, MA 02138, USA }
\author{A.~Yacoby}
\author{M.~D.~Lukin}
\affiliation{Department of Physics, Harvard University, Cambridge, MA 
02138, USA}

\date{\today}

\begin{abstract}
  We theoretically investigate the controlled dynamic polarization of
  lattice nuclear spins in GaAs double quantum dots containing two 
electrons. Three regimes of long-term dynamics are identified, including 
the build up  of a large difference in the Overhauser fields across the 
dots,  the saturation of  the nuclear polarization process associated with 
formation of so-called ``dark states,'' and the elimination of the 
difference field. We show that in the case of unequal dots, build up of 
difference fields generally accompanies the nuclear polarization   
process, whereas for nearly identical dots,  build up of difference fields 
competes with polarization saturation in dark states.  The elimination of 
the difference field does not, in general, correspond to a stable steady 
state of the polarization process.
\end{abstract}
\pacs{73.21.La, 76.60.-k, 76.70.Fz, 03.65.Yz}
\maketitle

\begin{figure}
\includegraphics[width=.45 \textwidth]{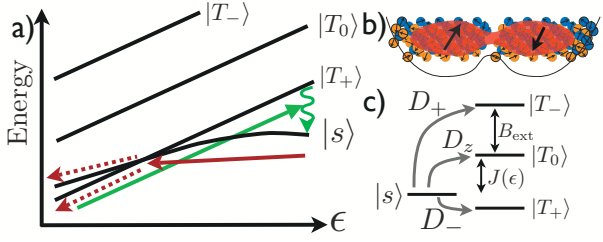}
\caption{\label{fig:qualfigabc} a) Two-electron energy levels as a 
function of detuning $\epsilon$ between $(1,1)$ and $(0,2)$ singlet 
states.  The  DNP cycle is illustrated by arrows.  b) A double quantum 
dot with two electrons interacting with a large number of lattice nuclear 
spins. c) Electronic energy level diagram with transitions from $s$ to triplet states $T
+,T_0,T_-$ driven by Overhauser fields $D_-,D_z,D_+$
respectively (gray arrows) and energies from external field $B_
\textrm{ext}$ and exchange splitting $J$ between $s$ and $T_0$ (black 
arrows).
 When $\bm{D}_\perp =0 $  electron-nuclear flip-flops are prevented, 
and when $\bm{D}=0$, electrons and nuclei decouple.  }
\end{figure}

Understanding the non-equilibrium quantum dynamics of localized
electronic spins interacting with a large number of nuclear spins is
an important goal in mesoscopic physics
\cite{Yusa05, Dixon97, Salis01, Ono04, Koppens08, Bracker05, Lai06}.
These interactions play a central role in spin-based implementations
of quantum information science, in that they determine the coherence
properties of electronic spin quantum bits \cite{Coish09}.
 One of the promising systems for realization of spin-based qubits 
involves electrically-gated pairs of quantum dots in GaAs, with one
electron in each quantum dot  (Fig.\ \ref{fig:qualfigabc}b) 
\cite{Hanson07}.   Hyperfine interactions with lattice nuclear spins are 
the leading mechanism for decoherence of the
electron spins,  and efforts are currently being directed towards
understanding these interactions \cite{Chen07,
Al-Hassanieh06,Christ07,Witzel08,Yao06,Coish04}, with the ultimate goal 
of turning the nuclear spins into a resource by controlling these 
interactions \cite{Klauser08,Petta10,Rudner09,Foletti09}.   Recent 
experiments have successfully demonstrated a wide variety of
electron-controlled nuclear spin polarization dynamics  
\cite{Foletti09,Petta08,Reilly08a,Foletti08}, but to date there is no 
unifying theoretical framework in which to understand the experimental 
results.

In this Letter we investigate theoretically the process of dynamic
nuclear polarization (DNP) in two-electron double quantum dots. This
process involves the preparation of the electronic spins in a singlet 
state and subsequent level crossing between the electronic singlet and 
triplet states with different projection of electronic angular momentum 
(Fig.\ \ref{fig:qualfigabc}a)  \cite{Petta08}. 
It is accompanied by nuclear spin flips, which polarize the spins of the
 nuclei inside the two dots, producing an 
effective magnetic (Overhauser) field for the electronic spins.  
Experiments demonstrate that DNP strongly modifies the difference 
between the Overhauser fields on the two dots, which is of central 
importance for control over singlet-triplet qubits 
\cite{Reilly08a,Foletti09}. 
Detailed understanding of DNP in these 
systems is both of fundamental interest and great practical 
importance for GaAs based electron spin qubits 
\cite{Ramon07,Ribeiro09,Yao09,Stopa10}.

In what follows we develop a theoretical framework to study the
non-equilibrium polarization dynamics of the nuclear spin environment.
Our approach takes advantage of the large effective temperature of the
nuclear spins and the short time-scale for electron spin evolution to
coarse grain the electronic system's dynamics, yielding a master
equation for the nuclear spin degrees of freedom, which we solve in a
semiclassical limit. Our key results may be understood by first 
considering three possible regimes that result from the DNP process. 
These include i) build-up of an effective difference field, ii) saturation in 
so-called ``dark states,'' and iii) preparation of nuclear spins in 
each quantum dot in states that produce identical Overhauser fields.

For example, \label{mark:heuristic} i) in the case of two dots with 
unequal sizes  the growth of an Overhauser difference field $D_z$ can 
be understood in the following heuristic picture, which is borne out by 
our analytic and numerical calculations.  Consider a system with a 
homogeneous wavefunction in the presence of both strong DNP pumping 
and nuclear dephasing.   The size difference results in different effective 
hyperfine interactions $g_{\ell(r)}$ on the left(right) dot.  
We find that the nuclear spins have nearly 
equal spin flip rates on the two dots, so that the build up of  the total 
Overhauser field $S_z$ is 
proportional to $g_\ell +g_r$, while the build up of $D_z$ is 
proportional to $g_\ell - g_r$.  Thus, $D_z$ tends to grow with $S_z$ 
such that $D_z/S_z \to (g_\ell-g_r)/(g_\ell+g_r)$. On the other hand, ii) when the dots are identical, 
or nearly so, we find a second regime at strong pumping, where $D_z$ 
does not grow and the polarization process shuts down the growth of $S_z$ 
by driving the difference field towards a dark state 
\cite{Taylor03}, with $D_x=D_y=0$. Such states are of interest for use as long-lived 
quantum memory. Finally, iii) electronic and nuclear degrees of 
freedom can be completely decoupled if two electrons are initially 
prepared in the singlet state, while the nuclear spins are prepared in 
a state with $\bm{D}=0$ (Fig.~\ref{fig:qualfigabc}c).  In such a
case, polarization stops and the dephasing time of the singlet-triplet 
qubit can be greatly extended.  However, we have not found physical 
parameter regimes in which such states can be {\it stably} prepared.

\emph{Model --} The hyperfine
coupling between a localized electron in dot $d=\ell,r$ (for
the left, right dot) and a nuclear spin $\bm{I}_{kd}$ at $\vec{r}_{kd}$, is 
given by $g_{kd}=
a_\txt{hf} \, v_0 \abs{\psi(\vec{r}_{kd})}^2$, where $\psi$ is the electron 
wavefunction, $v_0$ is the volume per nucleus, and $a_\txt{hf}$ is a 
coupling constant. The homogeneous limit is defined by $g_{kd}=g_d$ 
for all $k$. $\bm{S}$ and $\bm{D}$ are defined through the
collective nuclear spin operators denoting the Overhauser fields in
the left $(\vec{L})$ and right $(\vec{R})$ dots $\vec{L}= \sum_k
g_{k\ell} \vec{I}_{k\ell}$ and $\vec{R}= \sum_k g_{kr} \vec{I}_{kr}$ such that
 $\Svec=(\vec{L}+\vec{R})/2$, $\vec{D}= (\vec{L}-\vec{R})/2$.


For a double quantum dot with two electrons, we can
write the Hamiltonian for the lowest energy $(1,1)$ and $(0,2)$
electron states, where $(n,m)$ indicates $n$ ($m$) electrons in the
left (right) dot. In this subspace the effective Hamiltonian for the electron 
and nuclear spins takes the form $H=H_{el}+H_{\textrm{hf}} + H_{n}$, 
where 
\begin{align} \label{eqn:H}
H_{el}&=   \gamma_e \vec{\Bext} \cdot ( \vec{s}_\ell+\vec{s}_r )
+J(\epsilon) \vec{s}_\ell \cdot \vec{s}_r \nonumber\\
H_{\textrm{hf}}&=  \Svec \cdot (\vec{s}_\ell+ \vec{s}_r)+  \cos 
\theta(\epsilon)\, \Dvec \cdot (\vec{s}_\ell - \vec{s}_r )
\\ \nonumber
 H_{n}&=   -   \sum_{k,d}\gamma_n (\vec{\Bext} + \vec{h}_{kd} ) \cdot 
\vec{I}_{kd}  \nonumber
\end{align}
here $\bm{s}_{\ell(r)}$ is the electron spin in the left(right) dot,
$\gamma_e$ ($\gamma_n$) is the electron (nuclear) gyromagnetic 
ratio, where we consider spin $3/2$ nuclei of a single species, $
\vec{\Bext}=\Bext\zhat$ is the external magnetic field, $\cos
\theta(\epsilon)$ is the overlap of the adiabatic singlet state $\ket{s}$ with the 
$(1,1)$ singlet state as a function of the detuning $\epsilon$ between 
the $(1,1)$ and $(0,2)$ singlet states, and $J(\epsilon)$ is 
the splitting between $\ket{s}$ and $\ket{\TO}$  \cite{Taylor07}.  The rms
values of the components of $\vec{L},\vec{R}$ in the infinite 
temperature ensemble are $\Omega_{d}=(\sum_k g_{kd}^2 I(I+1)/
3)^{1/2}$. We define $\Omega = \sqrt{(\Omega_\ell^2+\Omega_r^2)/
2}\approx (10~\txt{ns})^{-1}$ for typical few-electron double dot experiments, and work in 
units where $\Omega=-\gamma_e=\hbar=1$. In addition to the nuclear 
Zeeman energy we include a ``noise'' term $\vec{h}_{kd}$, representing 
the fluctuating, local magnetic field felt by a nuclear spin at site $
\vec{r}_{kd}$, which could arise from e.g.\ nuclear 
dipole-dipole and electric quadrupole interactions.    
We estimate the scale of the fluctuations to be
such that a typical nuclear spin dephases at a rate of 
$1$-$50$ kHz \cite{Taylor07}.  

We find the nuclear
spin evolution semiclassically by treating the nuclei and electrons as 
mean fields when solving for the electron and nuclear dynamics, respectively.
This semiclassical approximation has been well studied in the context of central 
spin models and is generally reliable for extracting average quantities 
of high temperature, low polarization nuclear ensembles in dots with a
large number of nuclei $N$ (typically $\approx 10^6$ \cite{Taylor07}) \cite{Chen07,Al-Hassanieh06}.

Neglecting $H_n$, the nuclear spins evolve according to  $
\dot{\vec{I}}_{kd}=i[H_\text{hf},\vec{I}_{kd}]$, giving equations of motion
\be
\langle \dot{\vec{I}}_{kd}  \rangle = \frac{g_{kd}}{2} \big(  \mean{\vec{s}_
\ell +\vec{s}_r} \pm \cos \theta \mean{\vec{s}_\ell-\vec{s}_r} \big) \times 
\langle \vec{I}_{kd} \rangle
\ee
where the top sign applies for $d=\ell$.   We now replace $
\mean{\bm{I}_{kd}}$ with $\bm{I}_{kd}$ since we are treating the 
nuclear spins semiclassically.  Consider a pulse cycle $\epsilon(t)$ of 
duration $T\ll1/g_{kd}\approx\sqrt{N}/\Omega_d$.  In a single cycle we 
can average over the fast evolution of the electrons to arrive at the 
coarse-grained equations \cite{Christ07}
\begin{align} \label{eqn:SCeom}
\dot{\vec{I}}_{kd}(t)&\approx \frac{\vec{I}_{kd}(t+T) -\vec{I}_{kd}(t)}{T} = 
g_{kd}\, \vec{P}_d(t) \times \vec{I}_{kd}(t),\\ \label{eqn:Pddef}
\vec{P}_d (t)&=  \int_{t}^{T+t} \!\! \frac{dt'}{2T}[
\mean{\vec{s}_\ell +\vec{s}_r} \pm \cos \theta \mean{\vec{s}_\ell-
\vec{s}_r}],
\end{align}
where $\bm{P}_d$ is a slowly-varying, effective Knight magnetic field felt by the 
nuclear spins.

We now consider the class of pulse sequences employed in Refs.~
\cite{Reilly08a,Foletti09}, in which the electronic system is initialized in 
$\ket{s}$ at large $\epsilon$ and $\epsilon$ is swept slowly through the 
$\ket{s}$-$\ket{\Tp}$ resonance followed by a fast return to $(0,2)$ and 
reset of the electronic state via coupling to the leads.
(Fig.~\ref{fig:qualfigabc}a).  
This results in a build up of \emph{negative} 
polarization. For simplicity, we work in the limit where the electron spin 
flip probability per cycle is small and calculate $\vec{P}_d$ to lowest 
order in $\Omega/J$, $\Omega/\Bext$, $\Omega\, T$, and $\Omega /
\beta$, where $\beta^2=\frac{1}{2}\abs{dJ/dt}\lvert_{t=t_r}$ is the sweep 
rate at the resonance time $t_r$, \ie, $J[\epsilon(t_r)]=\Bext$.

To calculate $\mean{\bm{s}_d}$ we work in the Heisenberg picture.
Defining $\sigma_+^{m}= \ket{T_m} \bra{s}$, we can write
$(s_\ell^\pm- s_r^\pm)/2 = (\sigma_\pm^1 - \sigma_\mp^{-1} )/\sqrt{2}$ 
and $(s_\ell^z - s_r^z)/2=-(\sigma_+^0+\sigma_-^0)/2$.  Since $\Bext$, 
$J$, $\beta \gg \Omega$, we can set $\mean{\ket{T_n}\bra{T_m}}=0$ in 
$\langle d \sigma_+^{m'}/dt\rangle$ to obtain the first order corrections 
to the electronic state:
\begin{align} \label{eqn:sdot1}
\mean{\dot{\sigma}_+^0} & = -i \sqrt{2} v(t) D_z  + i J(t) \mean{\sigma_
+^0}, \\\label{eqn:sdot2}
\mean{\dot{\sigma}_+^{-1}} & = - i v(t) D_-  + i (J(t)+\Bext) \mean{\sigma_
+^{-1}},\\ \label{eqn:sdot3}
\mean{\dot{\sigma}_+^{1}} &= i v(t) D_+  + i (J(t)-\Bext) \mean{\sigma_
+^{1}},
\end{align}
where  $v(t) =\cos \theta(t) / \sqrt{2}$.
Since $J$, $\Bext\gg v\Omega$, Eqs.\ \ref{eqn:sdot1} and 
\ref{eqn:sdot2} can be adiabatically eliminated. To find $
\mean{\sigma^1_+}$, we formally integrate Eq.\ \ref{eqn:sdot3} and 
perform a saddle point expansion about the resonance time, assuming 
$v(t)$ is constant in this region, to reduce it to a Landau-Zener problem 
\cite{Gullans09a}. From this solution we calculate the average initial 
spin flip probability per cycle, $p_{f0} = 2 \pi v^2(t_r) \Omega^2 /
\beta^2$.

Putting these results into Eq.\ \ref{eqn:Pddef} gives
\be \label{eqn:Pd}
\vec{P}_d = \pm \big( \Gamma_0 \,\hat{z}\times \vec{D}_\perp- \Delta_0 
D_z  \hat{z} - \Delta_- \vec{D}_\perp  \big),
\ee
where $ \Gamma_0 = p_{f0}/  \Omega^2\,T$ arises from the polarization 
process via $T_+$, $\Delta_0=  \mean{2 v^2/J}_c$ and $\Delta_- =  
\mean{ v^2 /(J+\Bext)}_c$ arise from electron-nuclear exchange 
processes via the $T_0$ and $T_-$ states, respectively,
$\mean{\cdot}_c$ indicates an average taken over one cycle, and $
\vec{D}_\perp=D_x\hat{x}+D_y \hat{y}$. 

 Qualitatively, the effect of the $\Gamma_0$ term is to polarize the 
nuclear spins, but it also saturates the polarization by driving the 
nuclear spins into a dark state, $\bm{D}_\perp=0$.  
The $\Delta_0$ term drives the 
nuclear spins out of dark states, unless $D_z=0$ as well. 
Without noise, states with $\vec{D}=0$ 
are stationary during this DNP process; we refer to these as ``zero states."

\begin{figure}
\includegraphics[width=.45 \textwidth]{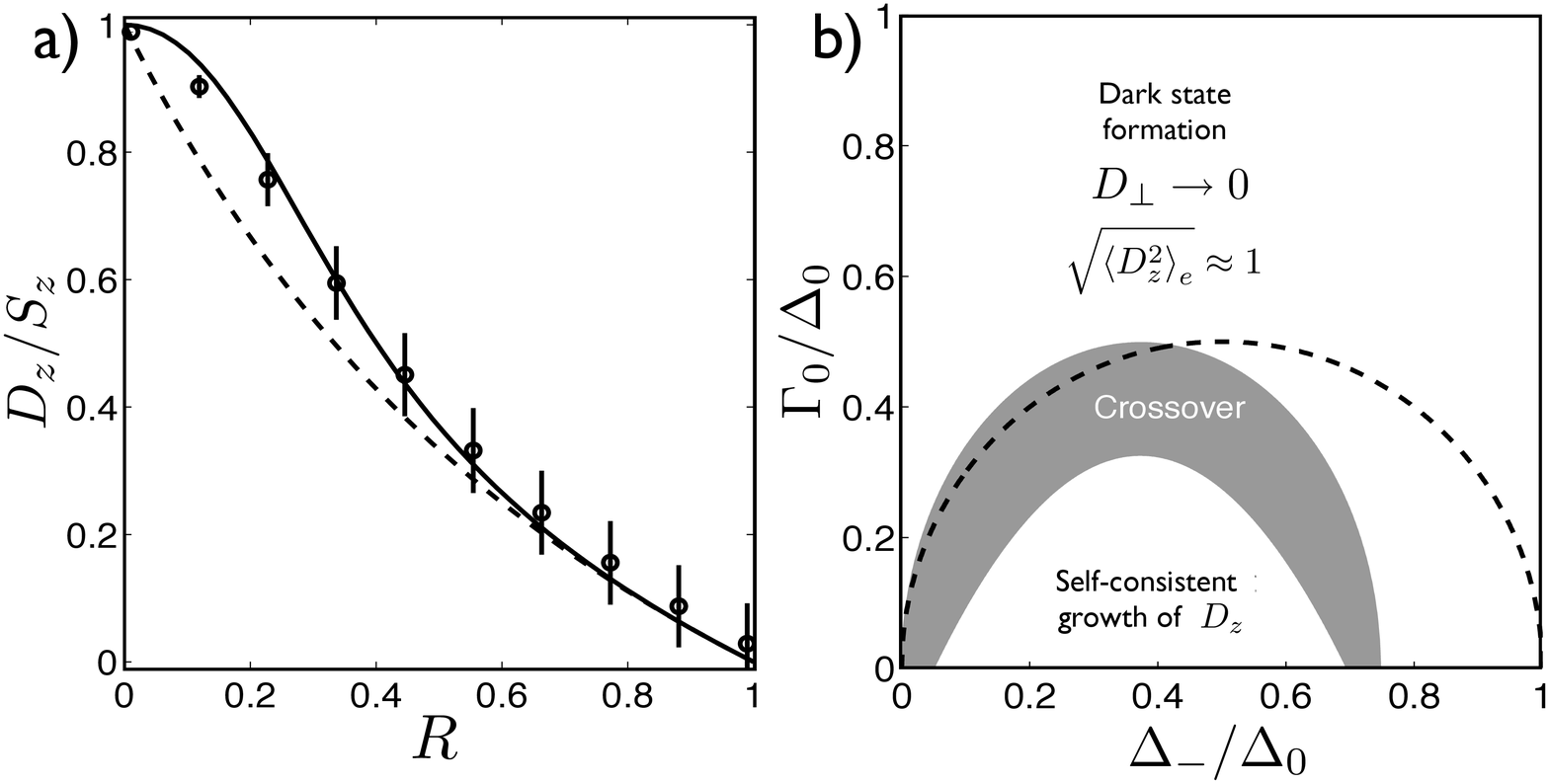}
\caption{\label{fig:DzSzRatio} 
a) Long time limit of $D_z/S_z$ as the relative hyperfine coupling in 
the two dots, $\rs= g_r/g_\ell$, is varied. The solid line is Eq.~
\ref{eqn:DzSzRatio} and the dashed line is $(1-\rs)/(1+\rs)$, obtained 
from a heuristic model (see text).  Circles are 
numerical results  with statistical error 
bars after averaging over an ensemble of 1000 initial conditions, run 
out to $t=10^5 /g_\ell \Gamma_0 \approx 1$ s, using an approximation 
to a 2D Gaussian electron wavefunction in terms of 100 coupling 
constants $g_{kd}$ with noise strength $\eta/g_\ell \Gamma_0 = 
10^{-3}$. b) Phase diagram for identical dots for either saturation in dark states 
or the self-consistent growth of difference fields as the DNP pumping rate (vertical axis)
and the Knight shift from $\ket{T_-}$ (horizontal axis) are varied relative 
to the Knight shift from $\ket{T_0}$.  The dark grey shaded region is a numerical
 ``crossover''  regime where both effects occur depending on initial conditions 
 and the dotted line is an analytic 
result from the simplified model of Eq.\ \ref{eqn:phasediagram}.  For typical polarization cycles 
$\Delta_- /\Delta_0 \approx 1/4$, but $\Gamma_0/\Delta_0 \approx 
p_{f0} B_\txt{ext}/\Omega^2 T$ can be tuned over a broad range.}
\end{figure}

Solving Eqs.~\ref{eqn:SCeom} with $\vec{P}_d$ given by Eq.~
\ref{eqn:Pd} for an arbitrary electron wave function is a challenging 
many-body problem. 
To help treat this problem, we have developed a new numerical method, 
which is formally equivalent to 
approximating the wave function by a unique set of $M\ll N$ coupling 
constants $g_{kd}$, that well approximates the time evolution of $
\vec{L}$ and $\vec{R}$ for a time that scales as $M$.  A full description 
of this method, which was used in Fig.\ \ref{fig:DzSzRatio}, along with a 
discussion of several higher order effects from finite magnetic field and 
adiabaticity, will be given elsewhere \cite{Gullans09a}.

\emph{Unequal dots --} Our results that zero states are unstable to
the growth of large difference fields, in the presence of asymmetry in the
 size of the dots and nuclear noise ($H_n$), can be shown analytically 
in the case of a simplified model.  We assume homogeneous coupling 
and work  in the high field, large $J$, limit where we can set $\Delta_0=
\Delta_- =0$ in $\vec{P}_d$. To treat the noise we first go into a frame 
rotating with the nuclear Larmor frequency, and assume $h_{kd}^{x,y}$ 
can be rotated away. 
We further assume that the nuclear noise can be approximated by a 
Gaussian, uncorrelated white noise spectrum,
$\gamma_n^2 \langle h_{kd}^z(t) h_{k'd'}^z(t')\rangle_n= 2 \eta\, \delta(t-
t') \delta_{kk'}\delta_{dd'}$, where
$\mean{\cdot}_n$ are averages over the noise \cite{Reilly08b}.
These local noise processes give rise to a mean decay of the collective 
nuclear spin variables $L_+$ $(R_+)$ and associated fluctuations $
\mathcal{F}_{\ell(r)}$, defined by $\mean{\mathcal{F}_d(t)\, 
\mathcal{F}_{d'}^*(t')}_n = 2 \Omega_d^2 \, \delta_{dd'} \delta(t-t')$.  As a 
result,
Eqs.~\ref{eqn:SCeom} and \ref{eqn:Pd}, including $H_n$, give
\begin{align}
\dot{L}_+&= g_\ell \Gamma_0 \, L_z (L_+ - R_+)/2 - \eta\, L_+ + \sqrt{2 
\eta} \, \mathcal{F}_\ell,\label{eqn:incohLp}\\
\dot{L}_z & = - \frac{g_\ell}{2} \Gamma_0\,  \big(L_\perp^2 -  \vec{R}_
\perp \cdot \vec{L}_\perp \big),\label{eqn:incohLz}
\end{align}
and similarly for $\vec{R}$.   From Eq.~\ref{eqn:incohLp}, if 
we start in a zero state, $\mathcal{F}_d$ will produce a fluctuation in $
\Dperp$, and the contribution to $\dot{L}_z$ of the form $- g_\ell 
\Gamma_0 L_\perp^2$ results, in the long time limit, in $L_z \ll -1$ and 
similarly for $R_z$.  Thus, $\vert\dot{L}_z/L_z \vert\ll 1$ and we can 
treat $L_z$, $R_z$ as static to find $\mean{L_\perp^2}_n$, $\mean{R_
\perp^2}_n$ and $\mean{\vec{L}_\perp \cdot \vec{R}_\perp }_n$, which 
allow us to find the slow evolution of $L_z$, $R_z$.  To lowest order in 
$1/S_z$ and $1-R$, where $R \equiv g_r/ g_\ell$,
\begin{equation} \label{eqn:incohGrad}
    \langle\dot{D}_z \rangle_n = -{\eta}   \left[  \mean{D_z}_n - (1- \rs) 
\mean{S_z}_n \right]/\mean{S_z}_n^2,
\end{equation}
and $\mean{S_z}_n = - \sqrt{\eta\, t}$.  This  growth of $S_z$ as 
$t^{1/2}$ is a result of our assumption of delta correlated nuclear noise.  
If  we assume a finite correlation time $\tau_c$ such that $
\mean{\mathcal{F}_d(t)\, \mathcal{F}_{d}^*(t')}_n =  \Omega_d^2 \exp(-
\abs{t-t'}/\tau_c)/\tau_c$, then for $g \Gamma_0 \abs{S_z} \ll 1/\tau_c$, $
\abs{S_z} \sim t^{1/2}$, but eventually $\abs{S_z} \sim t^{1/3}$.    
Integrating Eq.~\ref{eqn:incohGrad} gives $D_z/S_z \to (1-\rs)/2$.  For 
general $\rs$ we find, in the long time limit,
\be \label{eqn:DzSzRatio}
\frac{D_z}{S_z} \to \frac{1-R^2}{2\rs + \sqrt{4 \rs^2+(1-\rs)^4}}.
\ee
Fig.~\ref{fig:DzSzRatio}a shows good agreement between these results
and numerics for an \emph{inhomogeneous} Gaussian wavefunction.

\emph{Identical dots --}
For identical dots the previous arguments are no longer valid.  
Fig.\ \ref{fig:DzSzRatio}b, however, shows the results of numerical 
simulations \cite{Gullans09a} that demonstrate the existence of
a parameter regime for which there is self-consistent growth of
${D_z}$ even for identical dots.
%
Simulations were performed at each set of 
parameters by taking 20 different initially polarized nuclear spin 
configurations with $\Sz=-10$, $\Dz=-2$, $\eta/g_l\DelO$ between 
$10^{-2}-10^{-4}$, and a 2D Gaussian electron wavefunction 
approximated with 400 values of $g_{kd}$. We determined which 
parameter values had $\mean{\Dz}_e$ growing after $t=10^3/g_l
\DelO$. For $\GamO/\DelO>1/2$, no self-consistent growth of $\Dz$ 
appears, and the system approaches a dark state. For smaller $\GamO/
\DelO$ and for moderate $\Delm/\DelO$, continued growth of $\Dz$ 
is observed.  We find a similar boundary for unequal dots provided  $
\abs{1-R} \lesssim0.05$.

This phase diagram for identical dots can be verified analytically in
a simplified model, where the hyperfine coupling in each dot takes two 
values ($g_1\gg g_2$) on two groups of spins of similar size. We 
assume initially $-g_2 S_z \gg g_1 \abs{D_z} \gg 1\gg D_\perp$ with 
the polarization mostly in the strongly coupled spins. To lowest order in 
$g_2/g_1$, $\eta/g_2 D_z$, $g_1 D_z/g_2 S_z$, and $D_\perp/D_z$, 
we find \cite{Gullans09a}
\be \label{eqn:phasediagram}
\langle \dot{D}_z \rangle_n  \propto (\Gamma_0^2 + \Delta_-^2 - 
\Delta_0 \Delta_-) (g_1 \mean{D_z}_n/g_2 \mean{S_z}_n)^3.
\ee

Growth of $D_z$
requires nonzero $D_\perp$, but, as we show below, for large polarization
and weak noise $D_\perp \sim D_z/S_z$, 
which implies that the growth $D_z$ must occur self-consistently to prevent 
saturation.
This is illustrated by Eq.\ \ref{eqn:phasediagram}, where
the continued growth of $D_z$ is entirely determined by the sign of $
\Gamma_0^2 + \Delta_-^2 - \Delta_0 \Delta_- $.  For large $\Gamma_0$ or strong 
DNP pumping, 
the sign is positive, saturation effects dominate,
large difference fields are unstable and the system eventually reaches 
a dark state. For smaller $\Gamma_0$, the sign is negative and coherent
 evolution arising from interactions with 
$\ket{T_{0,-}}$ allows ${D_z}$ to continue
growing and $D_\perp$ remains finite.  Fig.\ \ref{fig:DzSzRatio}b shows 
reasonable agreement between this predicted boundary and our numerical 
results.

We now address the stability of the zero states in the absence of 
nuclear noise.  For identical dots, in the homogeneous limit, Eqs.~
\ref{eqn:SCeom} and \ref{eqn:Pd} give
\begin{align} \label{eqn:idEOM}
\dot{D}_+&= g\, i (\Delta_- - i \Gamma_0) S_z D_+ - g\, i \Delta_0 D_z  
S_+,\\
\dot{D}_z &= g \left[ (\Delta_- - i \Gamma_0) D_+ S_-  - c.c. \right]/2i. 
\label{eqn:cohDz}
\end{align}
Near a zero state $\bm{S}$ is constant since $\dot{\Svec} \sim O(D^2)$.  
The polarization, $g\GamO\Sz$, acts as a damping term for $D_+$; 
consequently, for $\Sz\ll-1$, $D_+ \to [\Delta_0 S_+/(\Delta_- - i 
\Gamma_0)]D_z/S_z $. Together with Eq.~\ref{eqn:cohDz} this implies 
$\dot{D}_z =0$.  Thus the stability matrix, $\partial \dot{D}_\mu/\partial 
D_\nu\vert_{\Dvec=0}$, has two negative eigenvalues 
  and one zero eigenvalue.
  Due to this zero eigenvalue, we expect
the stability of a zero state to be highly sensitive to external perturbations.  We 
find that inhomogeneous hyperfine coupling, 
multiple nuclear species, the hybridization of $\ket{s}$ and $\ket{T_0}$ 
as discussed in Refs.~\cite{Stopa10,Yao09}, and additional higher 
order corrections to $\vec{P}_d$ in $1/\Bext$ do not, however, break 
this zero eigenvalue.  In the absence of noise, we find numerically that 
for some parameters a large fraction of initial conditions result in the 
system spending a long time near a zero state; however, when we 
include nuclear noise or higher order corrections
in the inverse sweep rate, for example, zero states 
become repulsive on a long time scale \cite{Gullans09a}. Throughout this 
work we have mostly neglected nuclear spin diffusion \cite{Reilly08b} and 
spin-orbit coupling \cite{Rudner09}, both of which could potentially 
affect DNP and, in particular, the stability of zero states.

\acknowledgments
We thank S.\ Foletti, C.\ Barthel, M.\
Rudner, and I.\ Neder for valuable conversations. This work was 
supported by the Harvard-MIT CUA,
 the Fannie and John Hertz Foundation, Pappalardo, NSF,
the Physics Frontier Center, and the ARO.

\bibliography{DNP_DD}
\end{document}